Pseudogap and the chemical potential position topology in multiband superconductors

N. Kristoffel

Institute of Physics, University of Tartu, Ravila 14c, 50411 Tartu, Estonia

Original definition of the pseudogap [11,12] allowes to connect it conceptually with a topological insulator. It concerns the multiband superconductivity background with variable chemical potential position.

In the recent time the topological states of matter, e.g. [1-3], are extensively investigated. Topologically protected insulators and superconductors are known. The present letter analyses the conceptual connection of the pseudogap phenomenon [4-10] with a topologically protected band insulator. This protection is caused by the possible variable position of the chemical potential ($\mu$) in the multiband background of a superconductor. The multiband superconductivity is connected with multiple spectral gap order parameters. We exploit an original definition of the pseudogap [11,12] which exposes it as a possible natural event in the framework of multiband superconductivity. The motivation of our approach to the pseudogap problem, associated effects and experimental aspects can be found in papers [10,13-15]. In this letter we only stress that basing on our definition the pseudogap phenomenon can be associated with the topological organization of the active multiband background of the superconductivity. The topological nature of the pseudogap has been mentioned in the literature, e.g. [16-19], however on totally different basis as compared with the present one.

The designation "pseudogap" is predominantly used discussing low-energy excitations of high-temperature cuprate superconductors [4-9]. One observes here on the doping phase diagram a spectral gap evidently different from the superconducting gap. This "anomalous" gap exceeds markedly the superconducting dome described by the transition temperature ($T_c$) curve. It can be also detected inside of the superconducting region as observed in the normal state. The properties of these two type gaps demonstrate different behaviour with doping etc. In underdoped region the energy scales of them are also markedly different. Progressive doping washes out this dichotomy. At present there is a common conclusion that the superconducting gap ($\Delta$) and the pseudogap are of different genesis and compete (in some sense). An enormous amount of experimental and theoretical work has been done without a full understanding of the pseudogap and its origin. We associate the pseudogap with a specific excitation in the multiband superconductivity scheme.

Multiband superconductivity with appearing here interband pairing channels [20-24] has been known a long time ago. Such approaches have found numerous applications as being stimulated and followed by the discovery of new classes of materials. These compounds possess multicomponent Fermi surfaces and complex electron spectra. Multiband superconductivity opens various advantages and peculiarities as compared with the one-band BCS case.

For our approach to the pseudogap event it is of primary significance that on the multiband background the position of the chemical potential can vary between various bands or combinations of them. The Fermi surface momentum space region for location of μ becomes decisive including the possible Fermi surface reorganizations.

The chemical potential position can vary under given conditions with the developing electron spectrum "topology". There can be bands including μ, or not, in the full actual multiband complex. Eventual changes by doping and reorganizations of the electron spectrum rule this choice. The pairing mechanism can include essential contributions from the interband interactions. These interactions work effectively in the case of overlapping bands or small gaps between them. The doping process can include remarkable changes in the electron spectrum by changing the bands overlap [11-15]. This is a decisive circumstance for the behaviour of the pseudogap.

In multiband superconductivity the quasiparticle energies have the Bogolybov form

$$E_\alpha = \pm\sqrt{(\xi_\alpha - \mu)^2 + \Delta_\alpha^2}. \tag{1}$$

Here $\xi_\alpha$ is the band energy and $\Delta_\alpha$ the superconducting gap induced in the same band.

Our original definition [11, 12] of the pseudogap exposes it as the minimal quasiparticle excitation energy of the band not bearing μ. Hence the formation of a pseudogap is protected by the missing position of µ in the particular band. Correspondingly

$$E_\alpha(PG) = \sqrt{(\xi_\alpha - \mu)^2_{min} + \Delta_\alpha^2} \qquad (2)$$

with $|\xi_\alpha - \mu|_{min} \neq 0$.
For µ out of $\xi_\alpha$ the normal state contribution in this expression contains the generic normal state gap $|\xi_\alpha - \mu|_{min} \neq 0$. Here lies the origin of the pseudogap. It induces the corresponding insulating properties associated with this gap.

If the position of µ in the multiband spectrum can be considered as a topological property one can attribute the pseudogap to be built up on a topological band insulator. However, there is the superconducting contribution $\Delta_\alpha$ in E(PG). It can be also introduced by the interband interaction. The contribution of $\Delta_\alpha$ into E(PG) grows as $|\xi_\alpha - \mu|$ tends to its minimum. Nevertheless the normal state gap contribution included in the pseudogap definition allowes to consider it as being based on a topological band insulator. The pseudogap associates with the insulating behaviour of the corresponding region in the restricted momentum space. Experimentally the insulator to metal type transition accompanying the vanishing of the pseudogap is well known in the normal state [25,26]. Consequently one is really dealing with the insulating properties of the momentum space region occupied by the pseudogap excitations.

In the case where the reorganization of the spectrum allowes $|\xi_\alpha - \mu|_{min} = 0$ the pseudogap disappears. Further the E(PG) continues on the phase diagram as the superconducting gap $\Delta_\alpha$. This happens with entering of µ into this band.

The condition $|\xi_\alpha - \mu|_{min} = 0$ determines a quantum critical point. The µ position topology breaks here the protection of the presence of the pseudogap. A Lifshitz type transition in the electron spectrum takes place. One associates usually a quantum critical point with the vanishing pseudogap. The same result follows from our approach in accordance what one expects for topological transitions.

The reorganization of the Fermi surface at the critical point metallizes the carriers of the pseudogap spectral region. An insulator to metal transition accompanies the corresponding quantum critical point. In the superconducting state the critical point remains hidden in the $T_c$ dome at T = 0. The pseudogap compromised states fall off from the conductivity. Note that this can engange only a distinct part of the momentum space.

The pseudogap and the superconducting gap are different events. Both of them can be simultaneously characteristic to multiband superconductivity. Then not all the bands at the Fermi surface are intersected by µ. The $\Delta_\alpha$ and the $E_\alpha(PG)$ compete in the sense that from the pseudogap momentum space region an (essential) part of the spectral density is depleted from forming the superconductivity. There are situations where at the quantum critical point $T_c$ grows markedly by the contribution of liberated pseudogap states [27]. In the normal state at the quantum critical point the pseudogap band is gapless.

Because the interband pairing can be not very effective for the case where the gap $|\xi_\alpha - \mu|_{min} \neq 0$ is far from zero, another parallel pairing channel is expected to occur. Then the pseudogap $E_\alpha$ and a superconducting gap $\Delta_\beta$ can be simultaneously detected at the same doping level, but in different spectral windows. In the well exposed region of the pseudogap its slow temperature dependence stems from the superconducting contribution (2). The T*-line usually represented as the high-energy (temperature) limit for the presence of the pseudogap [4-9] will be destroyed by fluctuations [28], as also the corresponding insulating properties.

A three band model of cuprate superconductivity using the mentioned standpoints can be followed in [13-15,27]. The properties and associated events found in this approach agree qualitatively with experimental observations.

Our representation (2) for the pseudogap excitation enables it to be conceptually connected with a topological interpretation. This gap is essentially associated with a normal state topological band insulator. The latter is protected by the location of the chemical position in the manifold of multiband superconductivity background.

This work was partially supported by the European Union through the European Region Development Fund.